# On the Spin-Reorientation Transition in TmFeO$_3$

U. Staub[1], L. Rettig[1,§], E. M. Bothschafter[1], Y. W. Windsor[1,§], M. Ramakrishnan[1], S. R. Avula-Venkata[1], J. Dreiser[1], and C. Piamonteze[1]

[1]*Swiss Light Source, Paul Scherrer Institut, CH-5232 Villigen PSI, Switzerland*

V. Scagnoli[2,3]

[2]*Laboratory for Mesoscopic Systems, Department of Materials, ETH Zurich, 8093 Zurich, Switzerland.*

[3]*Laboratory for Multiscale Materials Experiments, Paul Scherrer Institut, 5232 Villigen PSI, Switzerland*

S. Mukherjee[4] and C. Niedermayer[4]

[4]*Laboratory for Neutron Scattering and Imaging, Paul Scherrer Institut, 5232 Villigen PSI, Switzerland*

M. Medarde[5] and E. Pomjakushina[5]

[5]*Laboratory for Scientific Developments and Novel Materials, Paul Scherrer Institut, 5235 Villigen PSI, Switzerland*

[§] *Current address: Department of Physical Chemistry, Fritz-Haber-Institut of the Max Planck Society, Faradayweg 4-6, Berlin 14915, Germany*

X-ray magnetic circular and linear dichroism (XMCD and XMLD) have been used to investigate the Fe magnetic response during the spin reorientation transition (SRT) in TmFeO$_3$. Comparing the Fe XMLD results with neutron diffraction and magnetization measurements on the same sample indicate that the SRT has an enhanced temperature range in the near surface region. This view is supported by complementary resonant soft x-ray diffraction experiments at the Tm $M_5$ edge. These measurements find an induced magnetic moment on the Tm sites, which is well-described by a dipolar mean field model originating from the Fe moments. Even though such a model can describe the 4*f* response in the experiments, it is insufficient to describe the SRT even when considering a change in the 4*f* anisotropy. Moreover, the results of the Fe XMCD shows a different temperature evolution through the SRT, which interpretation is hampered by additional spectral shape changes of the XCMD signal.

**I INTRODUCTION**

Understanding coupled antiferromagnetic (AFM) systems and their magnetic phase transitions is of fundamental interest in condensed matter physics. Transition metal perovskites with a general formula $RTO_3$ can accommodate magnetic ions at both the R and the T sites. Typically, the T-site is occupied by a 3*d* transition metal ion and the R site by a 4*f* rare-earth (RE) ion. Such configuration allows magnetic super-exchange interactions to exist between 3*d* transition metal ions as well as between 3*d* transition metal and magnetic RE ions. An archetypical example is the orthoferrite family of materials, whose magnetic ordering has been previously studied with neutron scattering.[1, 2] The magnetic structure of the Fe cage is well documented. [1],[3] It has been found that some of the $REFeO_3$ exhibit a spin-reorientation transition (SRT), at which the AFM easy axis rotates by 90 degrees when lowering the temperature. Due to the fact that the magnetic super-exchange interaction between the well-localized 4*f* states and the 3*d* ions is much weaker than between the 3*d* ions, the super-exchange between the 4*f* ions is usually neglected. Indeed its magnetic ordering temperatures are two orders of magnitude lower than that of the Fe sublattice.

However, the SRT occurs only in REFeO$_3$ perovskites in which RE is a magnetic 4$f$ ion, and the transition temperature varies dramatically for materials with different RE ions. For instance, Sm has the SRT above room temperature, whereas for Tm it is around 85K and for Yb around 10K. [4] This indicates that the magnetic state of the RE plays a role in the SRT, though very little is known about the role of the RE ions magnetic state in the vicinity of the transition.

Recent years have seen renewed interest in orthoferrites, as their magnetic SRT behavior may open new directions in the field of spintronics, with the goal of increasing the speed of magnetic recording well below the nanosecond regime. The focus has been on ultrafast manipulation of magnetic order, achieved by exciting the system with ultra short and intense optical pulses.[5-7] The goal of such an ultrafast switch is to increase the speed of magnetic recording well below the nanosecond regime, and it has indeed been shown that a significant spin reorientation can be obtained on ultrafast timescales in TmFeO$_3$. [5] Inducing the SRT with an ultrashort laser pulse also leads to coherent magnetic excitations represented by a coherent modulation of the magnetization. Even more interesting is that such magnetic excitation can be excited directly by momentum transfer from circular polarized optical pulses, which constitutes the first observation of the inverse Faraday effect.[6] More recently, it has been proposed that exciting two optical phonon modes with a controlled relative phase can mimic a magnetic field and result in an excitation of the spin system. [8]

TmFeO$_3$ crystallizes in Pbnm symmetry and orders antiferromagnetically far above room temperatures, containing four chemical formula units in the unit cell that is approximately $\sqrt{2}a_p \times \sqrt{2}a_p \times 2a_p$, with $a_\text{p}$ the cubic perovskite lattice constant. The spin structure is simple G-type with Fe moments antiferromagnetically ordered (staggered $M_{AF}$) pointing along the **a**-axis for temperatures above the SRT, whereas at temperatures below the SRT the moments point along the **c**-axis [1]. The SRT is characterized by an onset temperature $T_1$, at which the spins start to coherently rotate away from $M_{AF}$//**a**, and an end temperature $T_2$, at which all spins have reached their final direction $M_{AF}$//**c** (See fig. 1(a)). In addition to the simple AFM structure, there is a small spin canting caused by the Dzyaloshinskii-Moriya (DM) interaction that induces a weak ferromagnetic moment $M_F$. This moment rotates also coherently in the same a,c plane, for increasing temperatures from $M_F$//**a** to $M_F$//**c** through the SRT. An additional weak AFM spin canting is allowed by symmetry.[4] As it is in the order of 1.6%,[9] of the total moment we do not consider it further here. The strong dependence of the SRT on the 4f electron system, which itself does not magnetically order above 4K, has led to the general belief that the SRT is initiated by the strong 4f magnetic anisotropy associated to the large orbital magnetic moment and can vary strongly via thermal occupancy changes in the 4f orbital levels. As the $Fe^{3+}$ ion in the orthoferrites has a half filled shell that exhibits only a small magnetic single ion anisotropy, very small external forces such as the weak magnetic interaction between the 4f-3d shells can influence its spin easy axis.

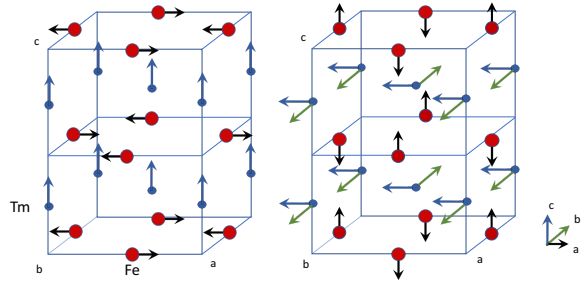

Fig. 1 Sketch of the magnetic structure of $TmFeO_3$ and the corresponding dipolar magnetic fields at the Tm site for right) $T>T_2$ and left) $T<T_1$.

Resonant x-ray techniques are powerful tools to investigate magnetic properties of materials,[10] in particular due to their element selectivity, which allows us to follow ferro- and AFM atomic moments for the different elements separately. For 3d transition metal ions these techniques are very sensitive in the soft x-ray regime and a number of different coupled magnetic transition metal oxide systems have been already studied in the past by both XMCD and resonant x-ray diffraction techniques.[11-15] In orthoferrites, XMCD and XMLD have been used to characterize the magnetic ordering phenomena of the Fe sublattice, or its interaction with other magnetic layers grown on it. [16-20]

In this paper, we discuss resonant x-ray diffraction at the Tm $M_5$ edge to study the magnetic structure and orbital orientation of the Tm ions through the SRT. Our results show that the $Fe^{3+}$ spins induce an antiferromagnetic component on the Tm ions only below the onset temperature of the SRT. This is compared to a dipolar mean field calculation. In addition, XMLD and XMCD at the Fe $L_{2,3}$ edges in reflection geometry are presented, which are sensitive to the antiferromagnetic (AFM) and ferromagnetic (FM) Fe components, respectively. These results show that the surface of our polished crystal has a much wider SRT than the bulk, which is visible in the XMLD of the Fe and Tm magnetic scattering, both probing the AFM component. The weak FM component exhibits a different temperature dependence indicating a decoupling of the FM and the AFM moments at the surface.

**II EXPERIMENTS**

Polycrystalline $TmFeO_3$ was prepared by a solid-state reaction. Starting materials of $Tm_2O_3$ and $Fe_2O_3$ with 99.99% purity were mixed and ground followed by heat treatment at 1000-1250$^0$C in air for duration of 70h with several intermediate grindings. Phase purity of the compound was checked with a conventional x-ray diffractometer. The resulting powder was hydrostatically pressed in the form of rods (7 mm in diameter and ~ 60 mm in length). The rods were subsequently sintered at $1300^0C$ for 20h.

The crystal growth was carried out using Optical Floating Zone Furnace (FZ-T-10000-H-IV-VP-PC, Crystal System Corp., Japan). The growth conditions were the following: growth rate was 5 mm/h, both rods (feeding and seeding rod) were rotated at about 20 rpm in opposite directions to secure the liquid homogeneity, 1.5 bar pressure of oxygen and argon mixture was applied during the growth process.

The crystal was oriented by Laue diffraction and different pieces cut along the [011] [100], and [001] directions. The surface was polished and annealed at $800^0C$ for 20h in oxygen flow. Magnetization measurements were conducted using a commercial MPMS SQUID magnetometer.

The resonant x-ray scattering experiments were conducted using the RESOXS ultrahigh vacuum diffraction end station [21] at the SIM beamline [22] of the Swiss Light Source (SLS). Linearly polarized incident light with either π or σ polarization (electric field in the scattering plane or perpendicular to it, respectively) were used for the resonant diffraction experiments of the (011) reflection at the Tm $M_5$ edge, and for the XMLD experiments at the Fe $L_{2,3}$ edges performed in reflection geometry. Circularly polarized light with opposite handedness was used for the XMCD experiments at the Fe $L_{2,3}$ edges in reflection geometry. For all measurements below the SRT, the sample was cooled through the SRT in a magnetic field of 0.1 T pointing along the a-axis to obtain a single magnetic domain state. This field was created by a permanent magnet that was removed before the measurements. Additional XMCD and XMLD experiments were performed at the X-Treme beamline [23] in total electron yield mode. For these experiments, the samples were covered with a 2-3 nm thick carbon layer to reduce charging, which could not be fully suppressed, limiting the data quality. Finally, neutron diffraction experiments were performed at the cold neutron triple-axis spectrometer RITA-II, SINQ, PSI using an incident wavelength of $\lambda = 4.21$Å from a pyrolytic graphite (002) monochromator and a 80' collimation between monochromator and sample. A pyrolytic graphite filter between monochromator and sample, a cooled Be filter between sample and analyzer reduced higher order contaminations of the incident beam. The sample was mounted in an orange cryostat with the (001) and (100) directions in the scattering plane.

**III RESULTS**

**a) Fe magnetic subsystems**

To study the FM Fe spin components, reflectivity spectra at 5 degree incidence were collected at 10K for opposite circular light polarization in the vicinity of the Fe $L_{2,3}$ edges. These are shown in Figure 2a) and 2b) for positive and negative field cooling through the SRT, respectively. Figures 2c and 2d show the corresponding XMCD response. Opposite XMCD signals are observed for opposite field cooling cycles indicating that the response is indeed magnetic in origin. A well-defined XMCD contrast is observed at the $L_2$ edge around 723.5 eV.

X-ray reflectivity spectra with $\pi$ and $\sigma$ incident polarization obtained at 10K, shown in Figure 3a and the corresponding XMLD contrast in Figure 3b, provide information about the AFM Fe order. Significant XMLD contrast is observed at both edges, as is expected from previous XMLD experiments on LaFeO$_3$.[16] Maximal contrast is observed at 711.5 eV, where further temperature dependence measurements were performed. No differences have been found for opposite magnetic field cooling within experimental accuracy, as expected for an antiferromagnetic XMLD signal (not shown).

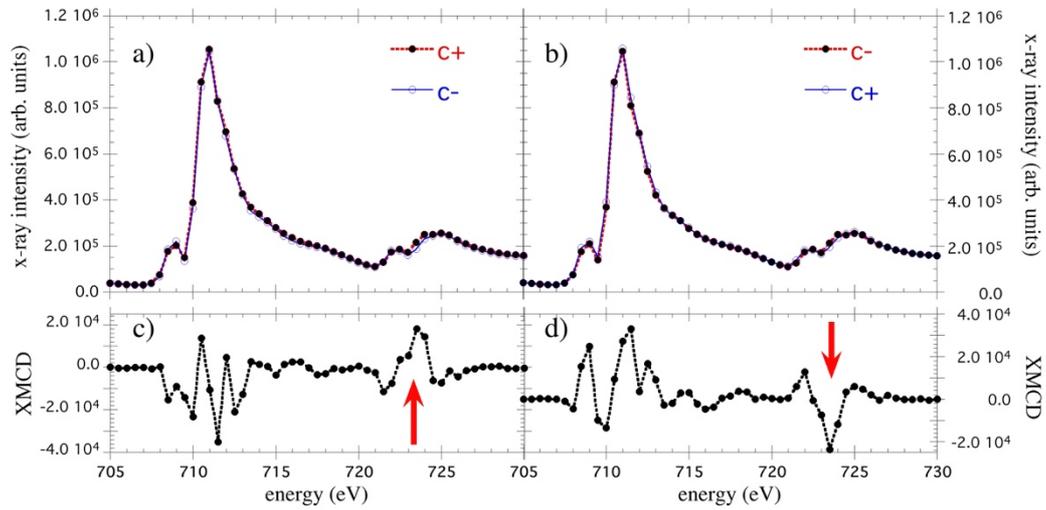

FIG. 2. Energy dependence of the x-ray reflectivity taken with opposite circular light polarization and its XMCD signals for positive field (H+) a), c) and negative (H-) field cooling, b), d).

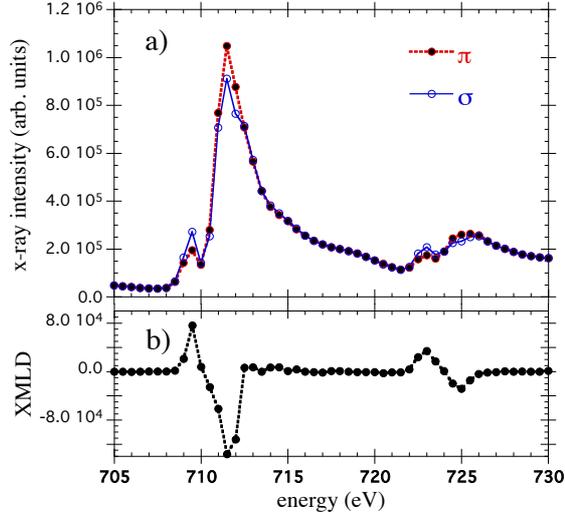

FIG. 3. Energy dependence of the x-ray reflectivity taken at five degree incidence angle for π and σ incident polarization a), and its corresponding XMLD signal b).

To study the Fe magnetic response through the SRT, the temperature dependent XMCD $(I_+-I_-)/(I_++I_-)$ and XMLD $(I_\sigma-I_\pi)/(I_\pi+I_\sigma)$ asymmetries were collected at the energies with optimal contrast. These are shown in Figure 4a and 4b. For increasing temperatures, the XMCD and XMLD asymmetries start to vary around the onset of the bulk SRT temperature ($T_1 \approx 81K$). The XMCD asymmetry shows an extremum around 100K. The XMLD shows a change of slope at $T_1$, however, it is followed by a linear temperature dependence up to approximately 120K, where a second kink indicates that the AFM contribution of the SRT transition is completed ($T_2$). The corresponding transition temperature $T_2$ from literature [4] ($T_1 \approx 91K$) is shown as a dotted line in Figure 4. Our results are clearly not compatible with it. This can be either due to fact that our $TmFeO_3$ crystal has different magnetic properties compared to previously published bulk samples or that the surface behaves differently. The probe depth of the XMCD and XMLD asymmetries taken in reflection geometry at these low incident angles is approximately 5 nm at resonance, making their response surface sensitive.

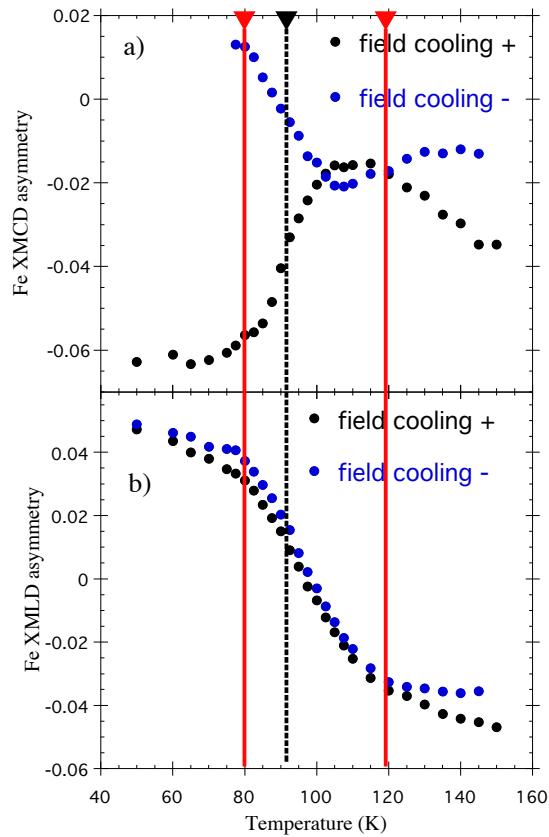

Figure 4. Temperature dependence of the XMCD and XMLD asymmetries measured in reflection for opposite field cooling. The red solid line shows lower and higher transition temperature $T_1$ and $T_2$ of the SRT, respectively, whereas the dashed black line shows the higher transition temperature for the published bulk data. [4]

To test the bulk magnetic properties, we performed magnetization measurements on the same crystal for magnetic fields along the (100) and (001) directions, which are shown in Figure 5. These data show that the bulk FM moments of the crystal exhibit $T_1$ and $T_2$ transition temperatures that are consistent with those reported in literature.

Note that any possible spin polarization at the Tm site at elevated temperatures is much weaker than the Fe moments, so the measured magnetization represents mostly the FM Fe spin canting. However, above bulk $T_2 \approx 92K$ there is still a slight increase in magnetization observed for increasing temperatures, peaking around 120K, at the same temperature at which the XMLD asymmetry changes slope. The AFM bulk properties were tested by collecting the neutron diffraction intensity from the (101) reflection in zero applied field. The intensity has both a structural and a magnetic scattering contribution. The AFM axis can either rotate through the [101] or [10-1] direction when rotating from the [001] to the [100] direction during the SRT. The temperature dependence, shown in Figure 6, allows us to distinguish between the two cases. It exhibits a distinct increase in neutron scattering signal through the SRT. This indicates that a majority of the domains rotate through the (10-1) direction, because the signal is sensitive to the magnetic moments perpendicular to the momentum transfer wave vector ($\mathbf{Q}$||(101)). Assuming that the temperature dependent intensity is caused by such a majority domain rotation and assuming for 50K $\mathbf{M}_{AF}$ // (001) and at 140K $\mathbf{M}_{AF}$ //(100), results in a smooth rotation of the spin direction, which is visualized in the inset of Figure 6. These results are consistent with the FM moments observed in the magnetization measurements, which indicate that the bulk FM moments remain pinned perpendicular to the AFM moments through the SRT. These bulk results clearly indicate that our $TmFeO_3$ crystal has the main magnetic properties as reported previously and that it is the surface regions that cause the difference in the XMCD and XMLD. The surface has a much higher $T_2$ and an extended SRT temperature range.

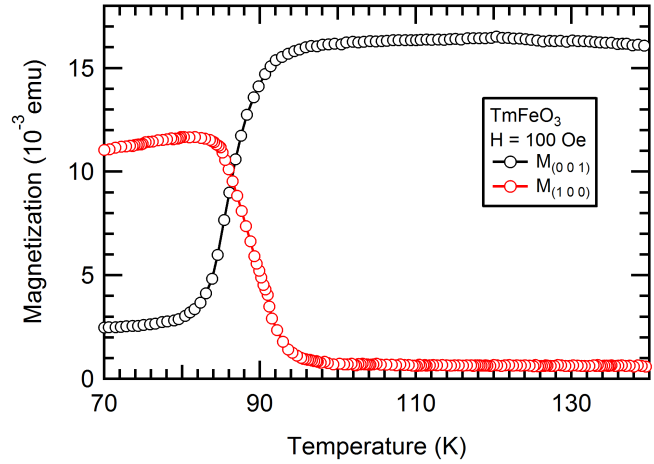

Figure 5 Temperature dependence of magnetization along the [001] and [100] directions of single crystal of $TmFeO_3$.

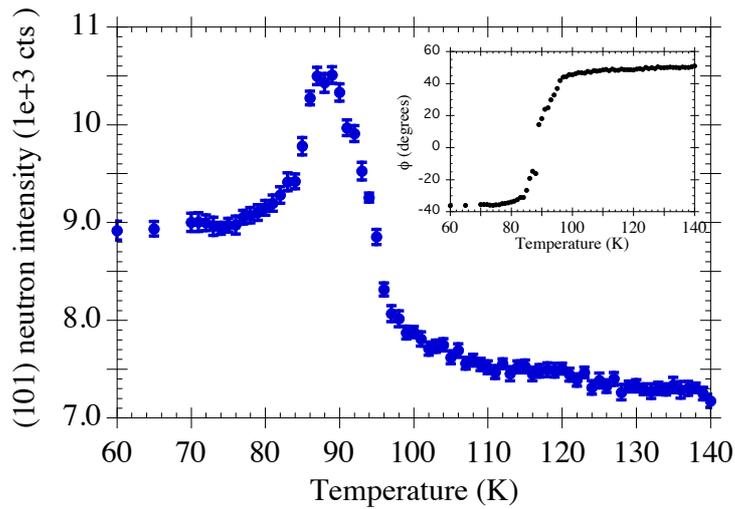

Figure 6 Temperature dependent neutron scattering intensity of the (101) reflection of $TmFeO_3$ containing both structural and magnetic components. Inset: Calculated rotation angle with respect to the momentum transfer (101) of the AFM Fe spin component extracted from the reflection intensities.

To confirm such a scenario, and in particular, to understand the unexpected behavior of the XMCD asymmetry taken in grazing incidence, standard XMCD measurements in total electron yield mode would be very helpful. However, due to the strong insulating character of the crystals, charging effects are too large to obtain reasonable XAS and XMCD spectra from the surface region in standard electron yield mode. To reduce the effect of charging, the sample was coated by 2-3 nm of carbon. Carbon coating of an oxide is relatively gentle and is expected not to significantly change the probed properties. Fig. 7 shows the XMCD spectra with the incoming x-ray beam along [100] taken from a (011) surface cut of the crystal. The sample was cooled in a magnetic field applied along the x-ray beam direction prior to the measurement. The XMCD data at the $L_3$ edge are still significantly affected by charging and only the data at the $L_2$ edge are sufficiently smooth to be dominated by the intrinsic XMCD. This is confirmed by the absence of a clear XMCD at 150K, as expected for when the FM Fe moment is expected to be close to perpendicular to the incoming x-ray direction.

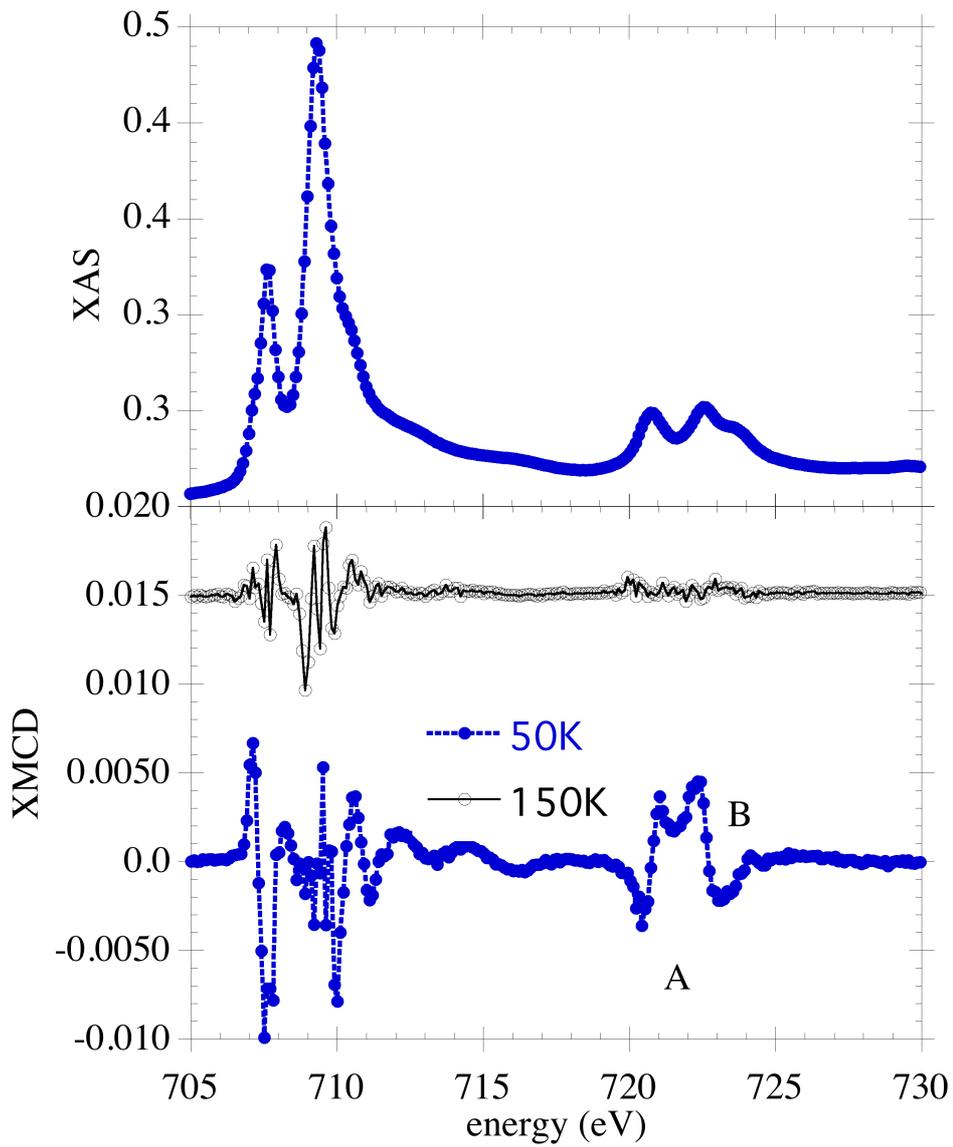

Figure 7 XAS (upper part) at T=50K and XMCD spectra (lower panel) of TmFeO$_3$ in the vicinity of the Fe $L_{2,3}$ edges at two different temperatures measured by total electron yield. The curves are offset for clarity. A and B represent the spectral features at which the temperature dependence has been taken.

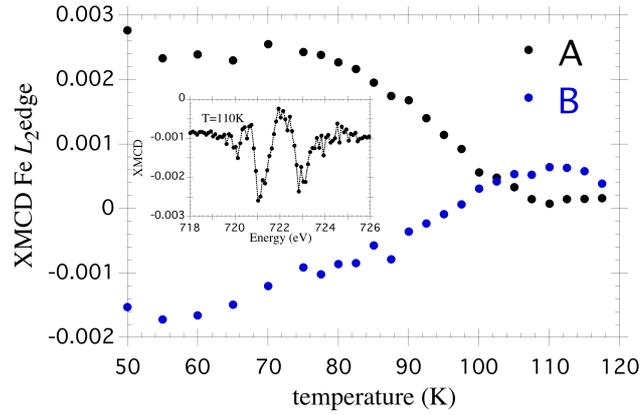

Figure 8: Temperature dependence of the averaged XMCD signal of spectral range A (720.8 to 722.6 eV) and B (722.7 to 724eV) measured by total electron yield. Inset: average XMCD temperatures 107.5-112.5 eV, showing a change of spectral shape compared to 50K (Fig. 7).

The temperature dependence of the average XMCD signal in region A (720.8 to 722.6 eV) and region B (722.7 to 724eV) are shown in Fig. 8. These data exhibit a crossing of the XMCD asymmetries at 103K and another possible cancelation at 120K. The same feature was observed in the temperature dependence of the reflectivity in Figure 4a. An average spectral shape is shown in the inset of the figure 8, which indicates that the crossing does not come from a cancelation of magnetic moments along the probe direction, but is rather due to a change of the spectral shape of the XMCD signal for different FM moment directions in the crystal.

**b) Tm 4*f*-Magnetism**

The energy dependence around the Tm $M_5$ edge of the (011) reflection for opposite field cooling is shown in Figure 9 for $\sigma$ and $\pi$ incoming linearly polarized x-rays. The sample was at an azimuthal angle $\Psi=180°$, which corresponds to the a-axis in the scattering plane. Thomson scattering is symmetry forbidden for the (011) reflection in Pbnm symmetry. However, the aspherical charge distributions of the resonant ions (Tm) have different orbital alignments at the four crystallographically equivalent 4c sites, which allows resonant scattering for this reflection. The individual Tm ions have different local crystal field axis orientations, which are directly related to the tilts of the oxygen octahedra. The resonant scattering cross section in the electric dipole-dipole approximation at the $M_5$ edge is sensitive to the antiferro-type ordered quadrupole electron density in the 4f shell. The temperature dependence of the expectation value of the ordered quadrupole(s) is dominated by the occupancy of the relevant low lying 4f states, which follow Boltzmann statistics. Tm 4$f$ quadrupoles have also been observed at the Tm $M_5$ edge in isostructural $TmMnO_3$, where the forbidden (010) reflection was investigated. [15, 24]

The energy scans of the (011) reflection taken at 10K for opposite field cooling differs significantly from each other (positive field cooling, Fig. 9a and negative field cooling Fig. 9b). This is also clearly visible in the corresponding asymmetry $(I_\pi-I_\sigma)/(I_\pi+I_\sigma)$, which shows maxima/minima just before the main $M_5$ resonance (Fig. 9c). This indicates that the diffraction intensity is not purely from the antiferro order of the 4f quadrupoles, but that it contains an additional magnetic contribution that is magnetic field dependent. At 100K, the (011) reflection is much weaker at resonance than at lower temperatures. This is a direct consequence of the depopulation of the 4$f$ ground state that reduces the asphericity of the 4$f$ electron density. In addition, the intensity difference for opposite magnetic field cooling is also strongly reduced resulting in strongly reduced asymmetry for linear polarization. It shows that the magnetic contribution has almost vanished at this temperature. A magnetic contribution is also observed when laterally scanning the sample surface in (011) Bragg condition. A spatially homogeneous magnetic signal is observed when the sample is field cooled (not shown), as expected in a single-domain state. A different result is expected in the presence of magnetic domains. Figure 10 presents such a scan at 40K, without prior field cooling. Clear contrast is observed between antiferromagnetic domains that are several 100 μm wide and up to 2-3 mm in length. This contrast disappears when heating the crystal to 170K, demonstrating the absence of a magnetic signal at this temperature.

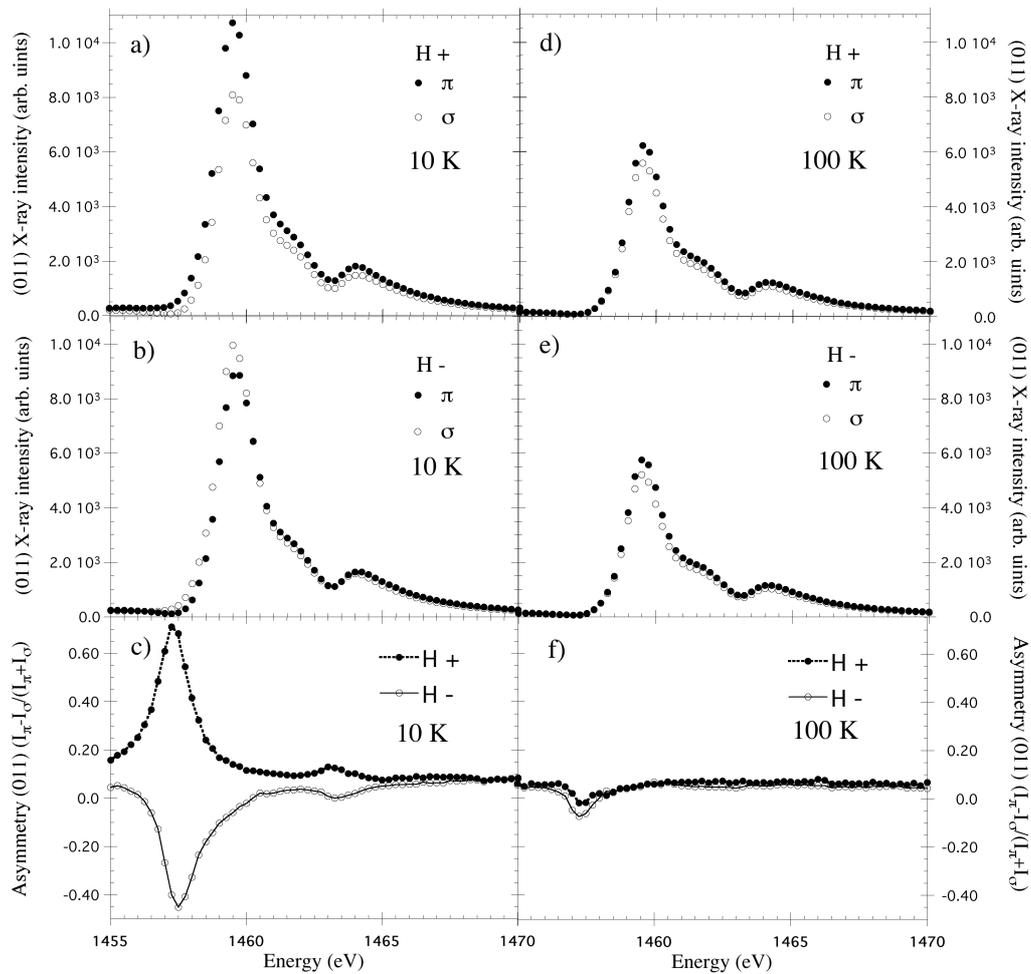

Fig. 9. Energy dependence of the (011) reflection in the vicinity of the Tm $M_5$ edge of TmFeO$_3$ for incident σ and π polarized x-rays for $\Psi$=180°, taken at T=10 K (a-c)) and 100K (d-e)), for opposite field cooling and its corresponding asymmetries. a) positive field cooling, b) for negative field cooling, c) corresponding asymmetry taken at 10K, d) for positive field cooling e) for negative field cooling, and f) corresponding asymmetry taken at 100K.

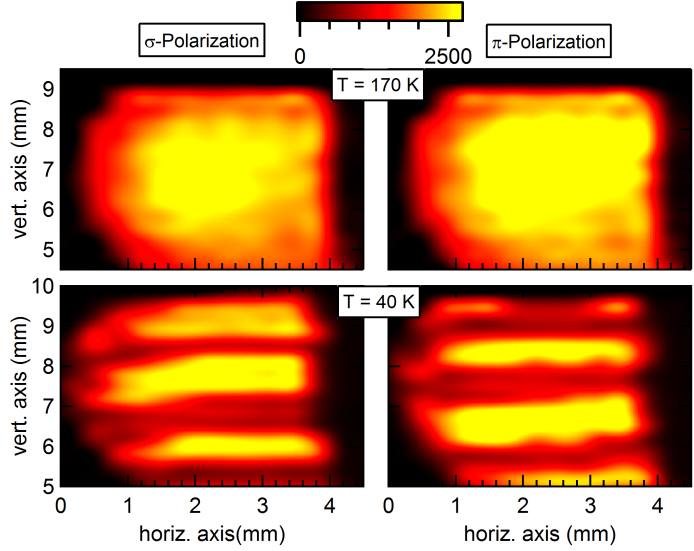

FIG 10. Intensity of the (011) reflection as a function of beam position on the sample, measured at 1457.2 eV (max. magnetic contrast) in the vicinity of the Tm $M_5$ edge of $TmFeO_3$ for incident σ and π polarization taken at T=40 K (lower panels) and 170K (upper panels). The data at T=40K are taken without field cooling and show the break up in opposite AFM domains.

To study the magnetic response of the Tm 4*f* system through the SRT, we follow the integrated intensity of the (011) reflection for increasing temperatures for σ and π polarization at the energy of the Tm $M_5$ edge with maximal magnetic contrast (1457 eV) and with maximal orbital (quadrupole) intensity (1459.2 eV). The intensity ratio between the two polarizations is shown in Figure 11 as a function of temperature for both energies and opposite field cooling directions. The magnetic contrast (solid symbols) strongly decreases upon warming, and is no longer detectable above

T≈120K, which coincides with $T_2$ obtained from the Fe XMLD signal. The magnetic contribution (open symbols) shows a much weaker contrast where the main orbital resonance energy. The temperature dependence of the magnetic signal shows a typical mean field like induced behavior for 4$f$ magnetic moments similar as found in NdNiO$_3$ [11] or TmMnO$_3$. [15] In isostructural TmMnO$_3$ the Tm ions are at the same crystallographic site with a similar crystal field potential. In addition, the Fe and Mn are both trivalent leading to a similar crystal field potential at the Tm site, allowing a comparison of spectral shapes between the two materials. It will be shown later that the same quadrupole is observed at the (010) and the (011) reflections in these systems resulting in an indeed similar spectral shape for these reflections. For the spectral shape of the magnetic scattering, the similarity is expected to be even larger, as the crystal field splitting of the 4$f$ states of a few ten's of meV is generally much smaller than the multiplet structure of a few eV at the $M_5$ edge. This leads also to very little variation in 4$f$ $M_5$ edges XMCD spectra for materials, which contain the same trivalent 4$f$ ions. The energy spectra reported in Ref. [15] for TmMnO$_3$ indicate that the magnetic scattering signal is maximal at a slightly lower energy than the maxima of the orbital scattering. This is in agreement with our data, and further supports the magnetic origin of the signal at 1457 eV in our TmFeO$_3$ system.

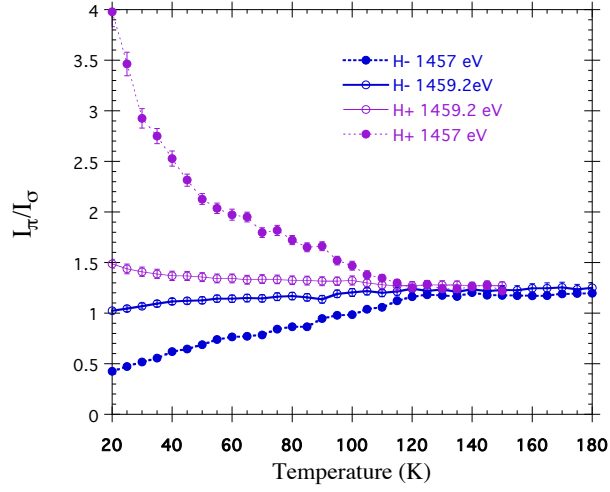

Figure 11: Temperature dependence of the (011) intensity ratio of σ and π polarization for cooling in opposite fields taken at the energy of maximum contrast (1457 eV) and at the maximum of the spectral shape intensity (1459.2 eV). All data are taken in the vicinity of the Tm $M_5$ edge.

To study the spatial symmetry of the quadrupolar order on Tm, we collected its azimuthal angle dependence (rotation about the scattering vector) in the low temperature phase at the maxima of the $M_5$ edge diffraction spectrum, which is shown in Figure 12.

Following Lovesey at al. [25], the resonant unit cell structure factor for the (011) reflection (with momentum **Q**) without the magnetic contributions can be written as

$$\Psi_Q^K = \sum_d e^{i\mathbf{Q}\mathbf{d}} \langle T_Q^2 \rangle_d \quad (1)$$

for which $\langle T_Q^2 \rangle$ is a Tm 4$f$ quadrupole with projection $Q$ (spherical coordinate system). The sum goes over all the Tm sites with individual positions **d**. Tm occupies the Wyckoff position 4c that has a ..m symmetry, which implies $\langle T_Q^2 \rangle = (-1)^Q \langle T_Q^2 \rangle$. This constrains $\langle T_Q^2 \rangle$ to be zero for Q=1 or -1 and it follows that the sum over the four Tm positions in the cell leads to

$$\Psi_2^2 = \Psi_{-2}^2 = 4i\cos(\gamma)\sin(2\pi y)\langle T_2^2 \rangle'' \tag{2a}$$

$$\Psi_1^2 = \Psi_{-1}^2 = 4i \sin(\gamma)\sin(2\pi y)\langle T_2^2 \rangle'' \tag{2b}$$

with $y$ the fractional atom position of the Tm ion and with $\gamma = atan(bl/ck)$, where $b$ and $c$ are the lattice constants, $k$, and $l$, the miller indices of the reflection. The unit cell structure factor has to be implemented into the scattering geometry to obtain the "global" structure factor and the corresponding intensity as a function of the azimuthal angle. For this we evaluate Eq. (B2) and (B3) of reference [26] and obtain only two non-zero elements of the quantity $A_1^2$ and $B_1^2$, from which we can obtain the structure factor for the different polarization channels [26]. It results in

$$F_{\sigma'-\sigma} = 4\sin(2\psi)\sin(\gamma)\sin(2\pi y)\langle T_2^2 \rangle'' \tag{3a}$$

$$F_{\pi'-\sigma} = 4[\cos(2\psi)\sin(\theta)\sin(\gamma) - \sin(\psi)\cos(\theta)\cos(\gamma)]\sin(2\pi y)\langle T_2^2 \rangle'' \tag{3b}$$

$$F_{\sigma'-\pi} = -4[\cos(2\psi)\sin(\theta)\sin(\gamma) + \sin(\psi)\cos(\theta)\cos(\gamma)]\sin(2\pi y)\langle T_2^2 \rangle'' \tag{3c}$$

$$F_{\pi'-\pi} = 4\sin(2\psi)\sin^2(\theta)\sin(\gamma)\sin(2\pi y)\langle T_2^2 \rangle'' \tag{3d}$$

with $\theta$ the Bragg angle of the (011) reflection. The corresponding azimuthal dependence is shown in Figure 11 and is in excellent agreement with the data. It shows that the reflection is indeed well described by a single 4f quadrupole contribution, with small deviations caused by the small magnetic contribution at this energy. Note that the calculation of the resonant scattering contribution of the space group forbidden (010) reflection results in the same single quadrupole component contribution, the rotated light channels is also proportional to the imaginary part of $\langle T_2^2 \rangle''$, with the absence of a scattering signal in the unrotated light polarization. This shows that indeed both reflections, the (010) and (011), are expected to have the same spectral shape.

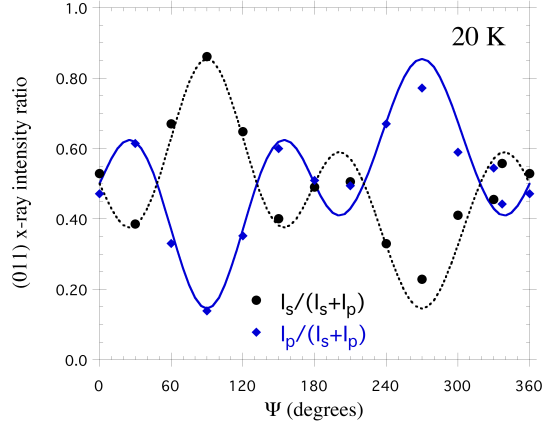

Figure 12: Azimuthal angle dependence of the intensity normalized σ and π polarized (011) reflection intensities taken at 20K at the maxima of the spectra. The lines are calculations as explained in the main text.

## IV DISCUSSION

The SRT in orthoferrites depends strongly on the 4*f* magnetic anisotropy. It is therefore interesting to study the interaction between the 4*f* and 3*d* moments that mediates the anisotropy between the two sublattices. The simplest approximation to describe the exchange interaction between the two magnetic subsystems is a mean field approach, in which the strongly coupled Fe moments lead to a net magnetic field at the paramagnetic Tm site. The very low magnetic ordering temperature of the Tm sublattice typical for oxides, reflect the very well localized 4*f* electron density and the corresponding weak super exchange interactions through the bridging oxygen, support such a simple assumption. As such, the effective field at the Tm site can be modeled using the magnetic dipole interaction, which successfully described the induced Tm moment that was extracted from azimuthal angle scans of resonant soft x-ray diffraction at several reflections in $TmMnO_3$.[15] Following such a promising starting point, we calculated the mean field created by the Fe moments at the Tm site as in Ref. [15]. Here we assume that the Fe moments are along the main crystallographic axis, as the canting from the weak Dzyaloshinskii–Moriya (DM) interaction is small. The resulting mean field from the Fe AFM moments along the **a** axis for $T>T_2$ produces a single FM component at the Tm site pointing along the **c** axis. This induced Tm moment enhances the FM spin component from the Fe canting. Note that also for $TmMnO_3$ the induced Tm moment was found to be perpendicular to the Mn moment, though the magnetic structure of the Mn sublattice is different. [15] The calculations for the magnetic phase with $T<T_1$ predict two different mean field components on the Tm, originating from the **c**-axis oriented AFM Fe moments: an AFM component along the **b** axis and a FM component along the **a** axis. The latter will further add to the Fe **a**-axis spin canting producing the overall bulk magnetization. This induced FM moment has the same magnitude as the one for $T>T_2$ (in comparison to the inducing Fe moment). The induced moment is expected to be larger in the low temperature phase as the single ion "paramagnetic" susceptibility is strongly temperature dependent. The AFM mean field component points along the **b** axis and is approximately 5 times larger than the FM component. The induced 4$f$ moment from this component will lead to a magnetic scattering signal at the (011) reflection, which is consistent with our observation.

The temperature dependence of the induced 4$f$ moment can be described by the temperature dependence of the effective field on the Tm site created by the Fe moments ($J_{3d-4f}$) and the response of the Tm ions to it. The response is described by the paramagnetic single ion susceptibility of the Tm ion along the effective mean field direction, which follows a Curie-Weiss like $1/T$ behavior over a wider temperature range. As the angle of rotation of the Fe spins from $\alpha=0$ to 90 is roughly linear between $T_1$ and $T_2$, (supported by XMLD data, neutron data and magnetization data) the field will be proportional to $\sin(\alpha)$ times the approximate $1/T$ behavior of the paramagnetic Tm single ion susceptibility. This gives a roughly linear intensity increase in temperature below $T_2$ that turns over to an approximately $1/T$ behavior below $T_1$, which is qualitatively observed in Figure 11.

To describe the SRT, we need to understand the reverse effect, the influence of the induced Tm moments on the Fe sites. Using again the dipole field approximation we calculate the effective field created by the Tm 4$f$ moments at the Fe sites ($J_{4f-3d}$). This results in field components that are parallel to the Fe AFM moments used as input in the previous calculations. This demonstrates that the Tm dipole mean field approximation is insufficient to describe the SRT, as it cannot mediate the 4$f$ anisotropy to the Fe sites. It becomes evident that even though a mean field dipolar field approach describes our data qualitatively correct, a direct or super exchange interaction between the 3$d$ and 4$f$ spin systems is required to explain the SRT, in addition to a variation in 4$f$ anisotropy.

An additional interesting point is the difference between the temperature dependence of the Fe XMLD and Fe XMCD signals (both taken in reflectivity and total electron yield mode). The Fe XMLD data exhibit an enlarged SRT temperature range. The XMCD data would be consistent with a rotation of the FM moment in a narrower temperature range (80-100K, which is still wider than the range for the bulk FM moment). This could be interpreted in terms of a decoupling of the FM spins from the AFM components, which would be puzzling. Such an interpretation would however require that the spectral XMCD shape remains constant during the SRT, which is not the case. Comparing the low XMCD spectra at 50K (fig. 7) ~ 110K (inset of figure 8), clear gradual changes are observed. This is consistent with moment rotation in magnetic fields of the Mn and Cr in XMCD measured at the $L_{2,3}$ edges in applied field. [27] These results show a clear change in spectra shape when rotating the magnetization. Therefore, an interpretation of the XMCD signal at a given energy in terms of rotation of the FM component would require a comparison and analysis of the spectral shape either in terms of sum rules or a comparison with first principle calculations, which goes beyond our current study.

**Conclusion**

Detailed resonant x-ray scattering and absorption experiments are presented on the Fe $L_{2,3}$ edges and the Tm $M_5$ edge of TmFeO$_3$ in the temperature range of the SRT transition. Clear XMCD and XMLD signals can be observed at the Fe edge, which allow us to separate the FM and AFM components of the Fe moments. Comparing these results with macroscopic magnetization and magnetic neutron diffraction intensities indicates that the SRT transition of the AFM component occurs continuously, but in a larger temperature range than reported in literature, suggesting that these results describe the magnetic response at the surface. It shows also that in the SRT temperature range the FM component is more complicated to analyze than the AFM component, as it is accompanied with change in spectral shape. An antiferromagnetic Tm moment is observed below $T_2$ that corroborate these findings. While the occurrence of the Tm spin polarization can be understood in terms of a dipole field approximation in a mean field approach, the dipolar interaction cannot explain the role of the Tm ions in the SRT. Our results indicate that the 3d-4f interaction has a significant non-dipolar contribution.

**Acknowledgements**

We acknowledge C.A.F. Vaz for carbon coating the sample and F. Nolting for fruitful discussions. We gratefully thank the X11MA, X07MA and X04SA beam line staff for experimental support. X-ray experiments where performed on the Swiss Light Source and neutron experiments were performed at the Swiss spallation neutron source SINQ, both at the Paul Scherrer Institut, Villigen, Switzerland. The financial support of PSI, the Swiss National Science Foundation and its National Center of Competence in Research, Molecular Ultrafast Science and Technology (NCCR MUST). E.M.B. acknowledges funding from the European Community's Seventh Framework Programme (FP7/2007- 2013) under grant agreement n.°290605 (PSI-FELLOW/COFUND).

References

[1]   J. A. Leake, G. Shirane, and J. P. Remeika, Solid State Communications **6**, 15 (1968).

[2]   S. M. Shapiro, J. D. Axe, and J. P. Remeika, Phys. Rev. B **10**, 2014 (1974).

[3]   H. Pinto, G. Shachar, H. Shaked, and S. Shtrikman, Physical Review B-Solid State **3**, 3861 (1971).

[4]   R. L. White, Journal of Applied Physics **40**, 1061 (1969).

[5]   A. V. Kimel, A. Kirilyuk, A. Tsvetkov, R. V. Pisarev, and T. Rasing, Nature **429**, 850 (2004).

[6]   A. V. Kimel, A. Kirilyuk, P. A. Usachev, R. V. Pisarev, A. M. Balbashov, and T. Rasing, Nature **435**, 655 (2005).

[7] S. Baierl, M. Hohenleutner, T. Kampfrath, A. K. Zvezdin, A. V. Kimel, R. Huber, and R. V. Mikhaylovskiy, Nature Photon. **10**, 715 (2016).

[8] T. F. Nova, A. Cartella, A. Cantaluppi, M. Forst, D. Bossini, R. V. Mikhaylovskiy, A. V. Kimel, R. Merlin, and A. Cavalleri, Nature Physics **13**, 132 (2017).

[9] V. P. Plakhty, Solid State Communications **47**, R1 (1983).

[10] S. W. Lovesey and S. P. Collins, *X-Ray Scattering and Absorption by Magnetic Materials* (Clarendon Press, Oxford, 1996), Vol. 1, Synchrotron Radiation.

[11] V. Scagnoli, U. Staub, Y. Bodenthin, M. Garcia-Fernandez, A. M. Mulders, G. I. Meijer, and G. Hammerl, Phys. Rev. B **77**, 115138 (2008).

[12] V. Scagnoli, U. Staub, A. M. Mulders, M. Janousch, G. I. Meijer, G. Hammerl, J. M. Tonnerre, and N. Stojic, Phys. Rev. B **73**, 100409(R) (2006).

[13] T. R. Forrest, S. R. Bland, S. Wilkins, H. C. Walker, T. A. W. Beale, P. D. Hatton, D. Prabhakaran, A. T. Boothroyd, D. Mannix, F. Yakhou, and D. F. McMorrow, J. Phys.: Condens. Matter **20**, 422205 (2008).

[14] S. B. Wilkins, T. R. Forrest, T. A. W. Beale, S. R. Bland, H. C. Walker, D. Mannix, F. Yakhou, D. Prabhakaran, A. T. Boothroyd, J. P. Hill, P. D. Hatton, and D. F. McMorrow, Phys. Rev. Lett. **103**, 207602 (2009).

[15] Y. W. Windsor, M. Ramakrishnan, L. Rettig, A. Alberca, E. M. Bothschafter, U. Staub, K. Shimamoto, Y. Hu, T. Lippert, and C. W. Schneider, Phys. Rev. B **91**, 235144 (2015).

[16] F. Nolting, A. Scholl, J. Stöhr, J. W. Seo, J. Fompeyrine, H. Siegwart, J.-P. Locket, S. Anders, J. Lüning, E. E. Fullerton, M. F. Toney, M. R. Scheinfein, and H. A. Padmore, Nature **405**, 767 (2000).

[17] L. Le Guyader, A. Kleibert, F. Nolting, L. Joly, P. M. Derlet, R. V. Pisarev, A. Kirilyuk, T. Rasing, and A. V. Kimel, Phys. Rev. B **87**, 054437 (2013).

[18] J. H. Lee, Y. K. Jeong, J. H. Park, M. A. Oak, H. M. Jang, J. Y. Son, and J. F. Scott, Phys. Rev. Lett. **107**, 117201 (2011).

[19] L. Joly, F. Nolting, A. V. Kimel, A. Kirilyuk, R. V. Pisarev, and T. Rasing, J. Phys.-Condens. Matter **21**, 446004 (2009).

[20] C. Y. Kuo, Y. Drees, M. T. Fernandez-Diaz, L. Zhao, L. Vasylechko, D. Sheptyakov, A. M. T. Bell, T. W. Pi, H. J. Lin, M. K. Wu, E. Pellegrin, S. M. Valvidares, Z. W. Li, P. Adler, A. Todorova, R. Kuchler, A. Steppke, L. H. Tjeng, Z. Hu, and A. C. Komarek, Phys. Rev. Lett. **113**, 217203 (2014).

[21] U. Staub, V. Scagnoli, Y. Bodenthin, M. García-Fernández, R. Wetter, A. M. Mulders, H. Grimmer, and M. Horisberger, J. Syn. Rad. **15**, 469 (2008).

[22] U. Flechsig, F. Nolting, A. Fraile-Rodrĺguez, J. Krempaský, C. Quitmann, T. Schmidt, S. Spielmann, and D. Zimoch, AIP Conf. Proc. **1234**, 319 (2010).

[23] C. Piamonteze, U. Flechsig, S. Rusponi, J. Dreiser, J. Heidler, M. Schmidt, R. Wetter, M. Calvi, T. Schmidt, H. Pruchova, J. Krempasky, C. Quitmann, H. Brune, and F. Nolting, J. Synch. Rad. **19**, 661 (2012).

[24] M. Garganourakis, Y. Bodenthin, R. A. De Souza, V. Scagnoli, A. Dönni, M. Tachibana, H. Kitazawa, E. Takayama-Muromachi, and U. Staub, Phys. Rev. B **86**, 054425 (2012).

[25] S. W. Lovesey, E. Balcar, K. S. Knight, and J. Fernández-Rodríguez, Phys. Rep. **411**, 233 (2005).

[26] V. Scagnoli and S. W. Lovesey, Phys. Rev. B **79**, 035111 (2009).

[27] G. van der Laan, R. V. Chopdekar, Y. Suzuki, and E. Arenholz, Phys. Rev. Lett. **105**, 067405 (2010).